# On-chip real-time detection of optical frequency variations with ultrahigh resolution using the sine-cosine encoder approach

X. Steve Yao[1,2,*], Yulong Yang[1], Xiaosong Ma[1], Zhongjin Lin[3], Yuntao Zhu[3], Wei Ke[4], Heyun Tan[3], Xichen Wang[5], and Xinlun Cai[3,*]

[1]Photonics Information Innovation Center and Hebei Provincial Center for Optical Sensing Innovations, College of Physics Science and Technology, Hebei University, Baoding 071002, China
[2]NuVision Photonics, Inc, Las Vegas, NV 89109, USA
[3]State Key Laboratory of Optoelectronic Materials and Technologies, School of Electronics and Information Technology, Sun Yat-sen University, Guangzhou 510275, China
[4]Liobate Technologies, Guangzhou, China
[5]NeoPIC Technologies, Suzhou, China

*Corresponding authors: X. Steve Yao syao@ieee.org and Xinlun Cai caixlun5@mail.sysu.edu.cn

## Abstract

Real-time measurement of optical frequency variations (OFVs) is crucial for various applications including laser frequency control, optical computing, and optical sensing. Traditional devices, though accurate, are often too large, slow and costly. Here we present a photonic integrated circuit (PIC) chip, utilizing the sine-cosine encoder principle, for high-speed and high-resolution real-time OFV measurement. Fabricated on a thin film lithium niobate (TFLN) platform, this chip-sized optical frequency detector (OFD) (5.5 mm × 2.7 mm) achieves a speed of up to 2500 THz/s and a resolution as fine as 2 MHz over a range exceeding 160 nm. Our robust algorithm overcomes the device imperfections and ensures precise quantification of OFV parameters. As a practical demonstration, the PIC OFD surpasses existing fiber Bragg grating (FBG) interrogators in sensitivity and speed for strain and vibration measurements. This work opens new avenues for on-chip OFV detection and offers significant potential for diverse applications involving OFV measurement.

## Introduction

Many optical distributed sensing applications, such as FMCW LiDAR and OFDR[1-3], require the measurement of fast optical frequency (wavelength) variations with high resolution for obtaining the k-clocks for data processing. Similarly, for FBG based quasi-distributed sensing applications, the ability for FBG interrogators to rapidly determine the wavelength variations of the reflected light from the sensing FBGs with high spectral resolution can greatly help to improve the performances of the system[4]. On the other hand, for optical computing involving micro-ring resonators (MRR) and wavelength division multiplexing (WDM)[5-8], the real-time detection of optical frequency variations (OFV) with high resolution is attractive for the accurate control of each WDM channel's frequency for precisely adjusting the weights of the WDM channels. In addition, the frequency control and monitoring of on-chip narrow linewidth lasers[9-12] is also important for their intended applications. Unfortunately, the measurement speed and resolution of the traditional optical frequency or wavelength measurement devices, such as wavemeters[13-17] and optical spectrum analyzers (OSAs)[18] are generally not sufficient to meet such demanding applications, although they have exceptionally high wavelength measurement accuracies. In addition, the large size and high cost of these devices generally prohibit them from being included in the sensing and optical computing systems.

To overcome these issues, unbalanced Mach-Zehnder interferometers (UMZI) are often used in the distributed sensor systems to measure the optical frequency for getting the required f-clock or k-clock[2,3]. The output signal of the UMZI is proportional to $\sin[\Delta\varphi(t) + \varphi_0]$, where $\Delta\varphi(t) = 2\pi\tau\Delta f(t)$ is the phase induced by the OFV $\Delta f(t)$, $\tau$ is the time delay corresponding to the optical path delay (OPD) between the two interferometer arms, and $\varphi_0$ is a constant phase. By measuring the phase $\Delta\varphi(t)$, the optical frequency variation (OFV) can be obtained, with a resolution inversely proportional to the time delay $\tau$. In practice, zero-crossings of the UMZI signal are used to determine the OFV with a resolution equaling to one half of the free spectral range (FSR) of the UMZI. Therefore, a very long length of optical fiber, up to few hundred or even thousand meters, must be used in the UMZI for the required OPD to get sufficient measurement resolution, which prevents the scheme being implemented on a photonic integrated circuit (PIC) chip. In addition, as discussed in[19], the long OPD enlarges the contribution of the laser frequency (or phase) noises in the interference signal, causing errors in the f-clock generations. Furthermore, the laser frequency scan ripples around the zero crossings of the UMZI output signal can also cause large frequency measurement errors[19,20]. Finally, the large OPD makes the speed of the interference signal corresponding to the OFV very high, which significantly increases the cost and power consumption of the electronic circuitry and therefore the cost of the detection system.

Hilbert transform of the UMZI output signal can be used to obtain the cosine function of the phase term $\Delta\varphi(t)$, which can be used together with the sine function of $\Delta\varphi(t)$ to determine the OFVs with much higher resolution and therefore can significantly reduce the required OPD in the UMZI[21]. Unfortunately, the process is not real time because a long length of $\Delta\varphi(t)$ data must be taken first before performing the Hilbert transform. Therefore, such a method is not suited for obtaining real-time k-clock or control signals in frequency stabilization and sensor applications that requires measuring the OFVs in real time, although it has been used for compensating the frequency scan nonlinearity of the tunable laser used in the system during post data processing[2,3,22].

To measure the OFVs in real time with high frequency resolution, several sine-cosine techniques have been proposed and demonstrated[19,23-25]. The basic idea is to simultaneously get the sine and cosine functions of $\Delta\varphi(t)$ induced by the OFVs, similar in principle to the sine-cosine encoder commonly used in electrical motors to determine the rotation angle increments[26,27]. In the implementation, only a very short OPD, on the order of 1 mm, is required for achieving an optical frequency resolution on the order of 10 MHz, sufficient for most applications[19]. Unfortunately, the demonstrated sine-cosine optical frequency detectors (OFD) were implemented with birefringence crystals or discrete optical components, which are too large to be integrated on a photonic chip.

Integrated chip-scale wavemeters have been successfully demonstrated using different types of unbalanced interferometers[28,29] similar in principle to the sine-cosine techniques, however, these devices are intended for high accuracy wavelength measurement and therefore most attention has been paid to the compensation of the thermal drift of the waveguides. Almost no efforts are made towards the detection of the detailed dynamics of the optical frequency variations with high speed and high resolution.

Venier optical frequency combs and microcombs have also been successfully demonstrated to measure fast OFVs with high speed and high accuracy[30-33], in which dual combs with slightly different repetition rates are used to beat with the CW laser under test. The instantaneous frequency of the laser can be obtained with high resolution, high accuracy and high speed by accurately measuring the beat frequencies. Frequency chirp up to 1500 THz/s with a repeatability of within 1 GHz can be achieved[31]. For a low chirp OFV, an absolute frequency can be calibrated to within 2.5 kHz with the dual comb vernier measurement approach. The major drawback of the methods relying on optical frequency combs is the high complexity and high cost, which is difficult in its present form to be embedded in sensor and measurement systems requiring low cost and compact size. Perhaps more fundamentally, this method is not real-time because it requires to take a large amount of data first to analyze before the optical frequencies can be obtained, making it difficult to be used for real-time feedback frequency control.

In this paper, we demonstrate a sine-cosine OFD implemented on a PIC chip for measuring rapid OFV, with a robust demodulation formulism to account for device imperfections in a wide wavelength range from 1480 to 1640 nm. In particular, the thin film lithium niobate (TFLN) platform is used to implement the OFD for convenience, specifically for the future integration with other TFLN based devices generating fast optical frequency sweeps on the same chip, such as tunable laser[34] and carrier suppressed single-side band modulator[35], in addition to other advantages TFLN offers[36,37], although other material platforms, such as SOI and SiN, can also be used. In addition, TFLN waveguide with a transmission loss of near 0.1 dB/cm has been demonstrated[38]. The resulting sine-cosine OFD chip measures only 5.5 mm x 2.7 mm, which can detect OFV with a demonstrated resolution down to 2 MHz (0.016 pm) and a speed up to 2500 THz/s, limited by the noise and speed of our electronics but already sufficient for most applications. We show that the OFV resolution on the order of 0.5 MHz with a speed of $3 \times 10^4$ THz/s can be achieved with our fabricated OFD chip using commercially available electronics of 1 MHz detection bandwidth. In addition, the required speed of the electronic circuit is reduced thousands of times compared to the UMZM approach described previously, resulting in large cost savings. We show that with this PIC OFD, tunable lasers' frequency scanning waveforms, nonlinearity, irregularity, speed, range, direction, and even tuning ripples not detectable with any other measurement methods can be clearly determined. Finally, we demonstrate that this PIC OFD can be used as a FBG interrogator with a strain measurement resolution of 0.1~0.2 με at 500 Hz. A strain resolution of 0.013 με and a temperature resolution of 0.0015 ℃ at a measurement bandwidth of 5 Hz are expected from the achievable OFV resolution of 2 MHz (0.016 pm), which is about 20 times better than the best FBG interrogators on the market. Pairing better commercially available electronics with the PIC OFD, 80 times better resolutions than those of the best commercial FBG interrogators can also be achieved. Our sine-cosine PIC OFD with the associated demodulation algorithm validates a new approach for the sensing PIC design and can be incorporated in different chip scale interrogators for various distributed and quasi-distributed sensing applications, as well as for on-chip laser frequency control and stabilization[39], which can be used for microring resonator (MRR) based optical computing schemes[8], among others.

## Results

### *Scheme and principle*

Fig. 1a illustrates the configuration of the proposed sine-cosine OFD. The light from a tunable laser with rapid frequency variation, as shown in Fig. 1b, is first split into two branches by a 50% coupler, one of which goes to an upper unbalanced interferometer (the main interferometer) and the other goes to a lower unbalanced interferometer (the assistive interferometer). Each interferometer consists of a 1×2 coupler and a 90° hybrid with two input ports and four output ports[40], with the upper one having a large OPD (or small FSR) between the two arms for high resolution frequency increment measurement, while the lower one having a small OPD, corresponding to a very large FSR, to ensure the output signal is monotonic in the whole wavelength (or frequency) measurement range of the OFD for estimating the absolute wavelength of light, as shown in Fig. 1c.

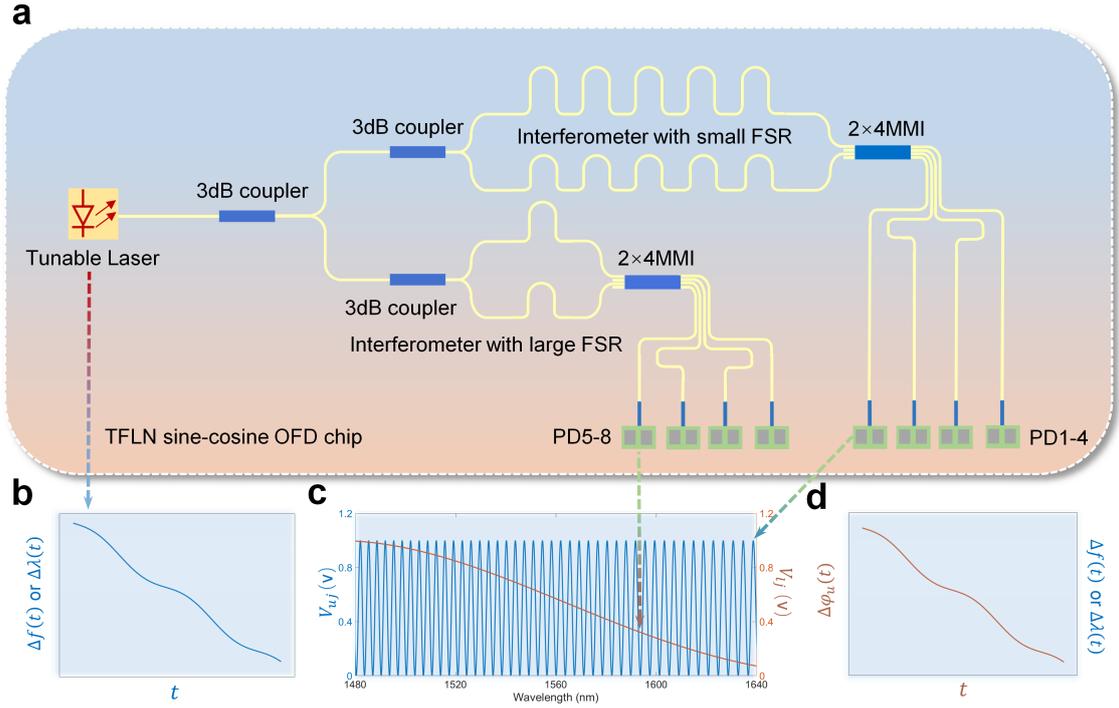

**Fig. 1. Configuration and measurement mechanism of the sine-cosine OFD. a** The schematic drawing of the sine-cosine optical frequency detector (OFD) consisting of two unbalanced I-Q interferometers with different free spectral ranges (FSRs) (TFLN: thin film lithium niobate, MMI: multimode interference coupler, PD: photodetector). **b** Frequency or wavelength of a tunable laser varies rapidly with time. **c** Illustration of the signal output from one of the PD in the upper (main) interferometer with a small FSR (blue) for high resolution OFV detection, and one of the PD in the lower (assistive) interferometer with a large FSR (red) for low resolution absolution frequency detection. **d** Measured $\Delta\varphi_u(t)$ from Eqs. (S7) to (S9) and the corresponding demodulated frequency or wavelength.

*Device calibration*

   Many parameters of the fabricated device are wavelength dependent and need to be calibrated. The setup for the calibration and evaluation of the PIC OFD is shown in Fig. 2, in which a laser with wide wavelength tuning capability is used as the light source to input into the OFD chip via a polarization controller to align the input polarization to that of the TE mode of the chip. The eight light outputs of the chip are detected by the corresponding PDs and are converted to eight voltages by two 4-channel transimpedance amplifiers (TIA, Koheron TIA 400-2k, 2 MHz bandwidth, 4 channels each) before being digitized by an 8-channel DAQ card (60 kS/s with 16-bit resolution) and sent to a personal computer. The details of the calibration process and the results are included in Supplementary Information Section 2 and 3.

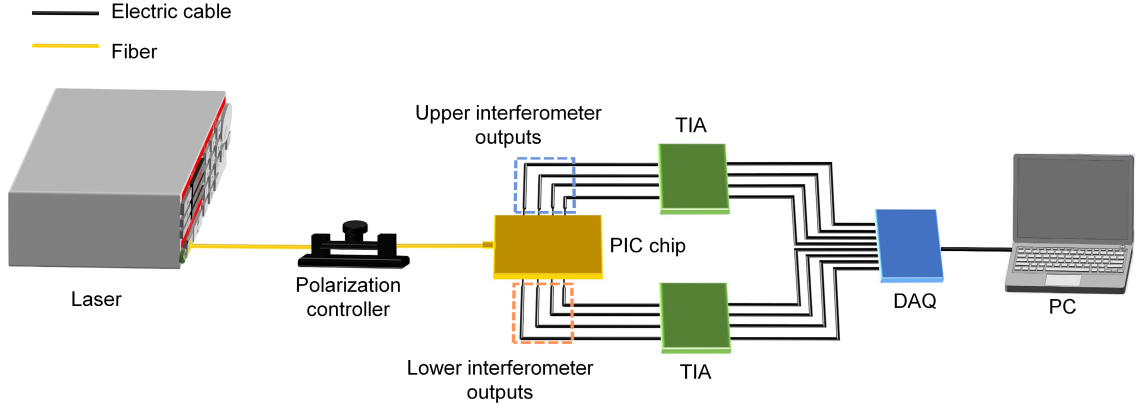

**Fig. 2. Setup for the calibration and evaluation of our PIC OFD.** PIC: photonic integrated circuit, OFD: optical frequency detector, TIA: transimpedance amplifier, DAQ: data acquisition card, PC: personal computer.

*Complete characterization of wavelength scan parameters of a tunable laser*

Fig. 3a shows the instantaneous wavelength of a tunable laser (Santec TSL-570) undergoing periodic scan (setting at 100 nm/s) in a range from 1500 nm to 1630 nm, measured with our PIC OFD after the calibration described in the previous section is performed. Clearly, the directions of the wavelength (frequency) variations can be unmistakenly identified, with the up-ramp and down-ramp of the linear wavelength scan clearly shown, which can be fitted to $\lambda_{up} = -0.72t^2 + 96.63t + 1380$ and $\lambda_{down} = -9.13t^2 - 146.08t + 2101$, both having an excellent goodness of fit $R^2 = 0.99999$. The scan periodicity $t_1$, and the time windows for the up and down ramps $t_2$ and $t_3$ are obtained to be 2.394 s, 1.43 s, and 0.68 s, respectively. The coefficient of the linear term represents the linear scan rate of the wavelength, showing that the down ramp is notably faster than the up ramp (191.18 nm/s vs. 90.91 nm/s). The scan rates in different time periods are shown in Fig. 3b, which are obtained by taking the derivative of the data in Fig. 3a. Note that there exist slight wavelength scan nonlinearities in the up and down ramps, represented by the coefficients of the quadratic terms of the fitted formula above, with the up ramp about 10 times smaller than that of the down ramp (-0.72 nm/s² vs. -9.13 nm/s²), which can also be seen in Fig. 3b.

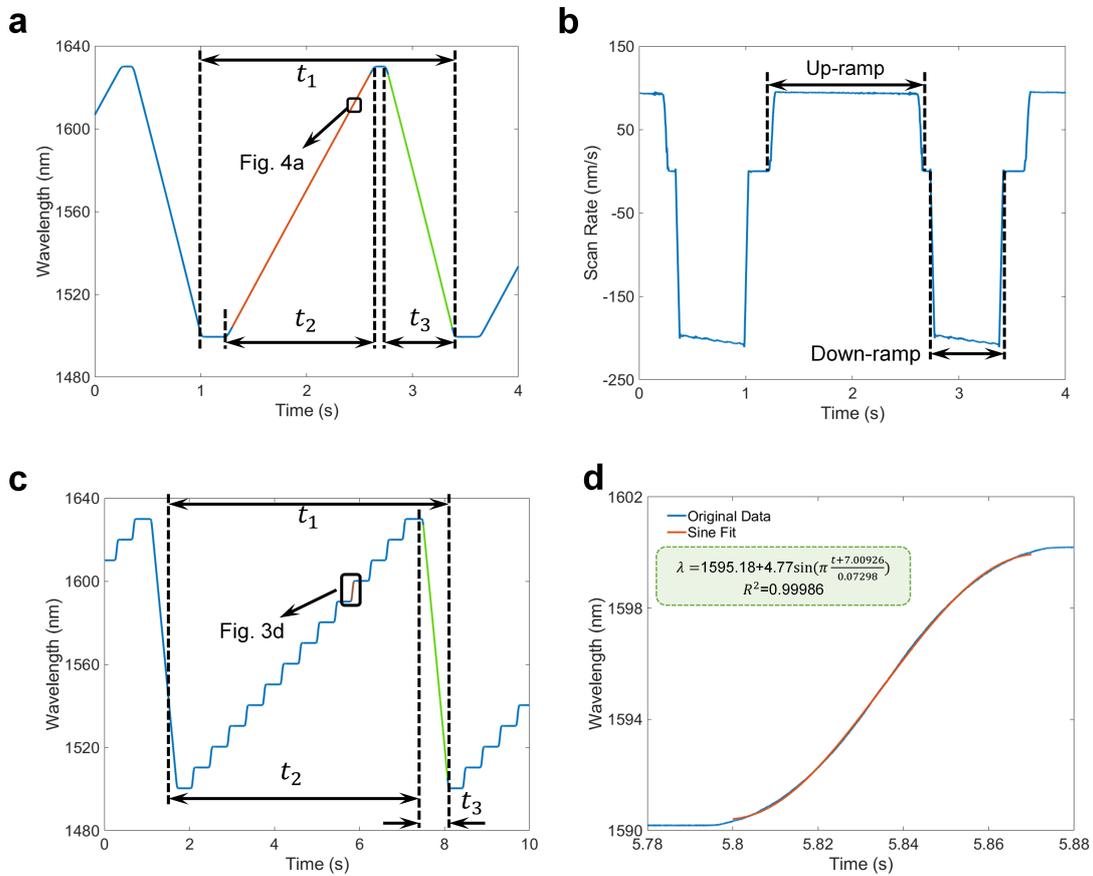

**Fig. 3. Characterization of wavelength scan parameters of a tunable laser. a** OFD measurement results of the instantaneous wavelength of a tunable laser (Santec TSL-570) being scanned in a range from 1500 nm to 1630 nm having a period of 3.5 s, with the down ramp (green) notably faster than the up ramp (red). **b** the corresponding wavelength changing rates in different time windows. **c** The OFD measurement results of the same laser being step-tuned, with the wavelength steps clearly shown. **d** the expanded view of a single step, showing the wavelength transition following a sinusoidal function. A low cost DAQ card (60 kS/s and 16-bit resolution) is used in the measurements.

Fig. 3c shows the measurement result when the laser is step-tuned at 10 nm per step, with the detail of each step clearly captured. The expanded view of each step is shown in Fig. 3d, with the transition between two steps following a sinusoidal function.

*Characterization of wavelength scan ripples and repeatability*

Because of the high wavelength (frequency) resolution of our PIC OFD, the details of the wavelength variations of a tunable laser can now be seen, which otherwise are not even known to exist by the laser manufacturers. Fig. 4a shows the zoom-in view of a small section of measured wavelength scan shown in the small box of Fig. 3a, in which small wavelength scanning ripples can be clearly seen. To rule out the possibility that the ripples are due to the measurement uncertainties, three repeated measurements were taken in three consecutive cycles of the wavelength scan. The highly repeatable results indicate that these ripples are indeed from laser scanning, not the measurement artifact. To further show the details of the scanning ripples, the deviations of the instantaneous wavelength from the linear fit curve of the wavelength up ramp are calculated and shown in Fig. 4b, which indicate that the ripples have a certain periodic pattern with a large periodicity of 4.8 ms and sub-periodicity of 0.93 ms, possibly due to the imperfections of the wavelength tuning mechanism, such as the motor or gears used for wavelength tuning.

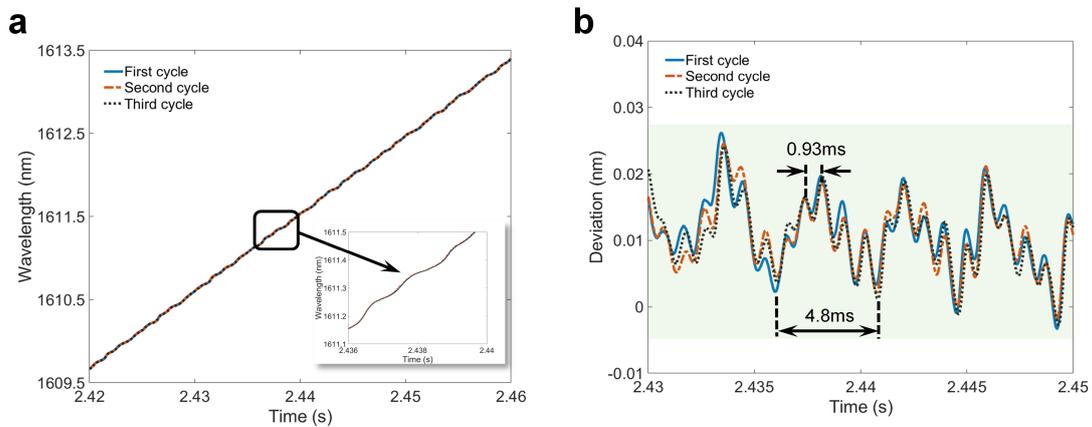

**Fig. 4. Characterization of wavelength scanning ripple and repeatability. a** Measurement data showing the wavelength scanning ripple, with three repeated measurements in three consecutive wavelength scanning cycles. **b** The wavelength deviations of three repeated measurements from a linear fit, showing the details of the wavelength scanning ripples.

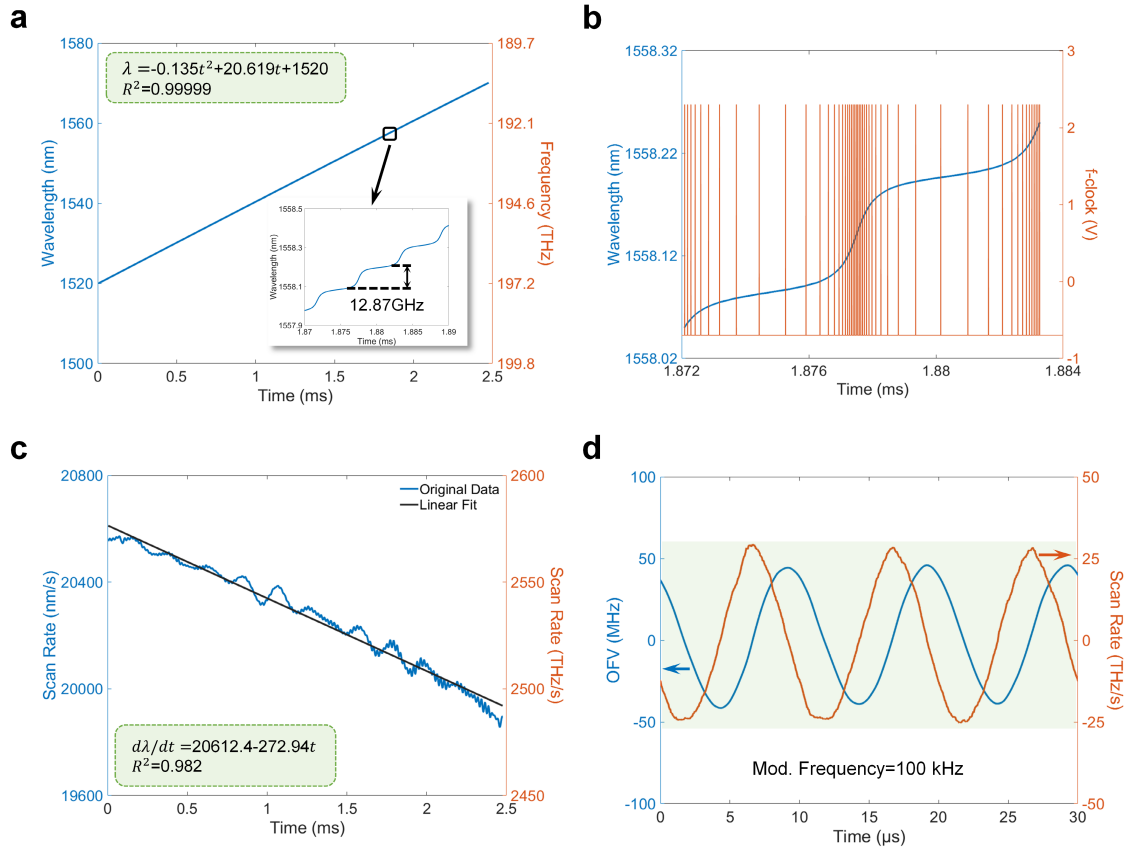

**Fig. 5. High speed frequency scan and modulation measurement results. a** Measured instantaneous wavelength (frequency) of a high-speed tunable laser (Newport TLB-8800-HSH-CL) scanned at a rate of 20000 nm/s (~2500 THz/s). Inset: the zoom-in view of the data showing stairway-like frequency increments. **b** The generated f-clock (k-clock, red line) with a frequency increment of 0.5 GHz and a pulse width of 10 ns. **c** The scan rate derived from a, with minor irregularities clearly shown. **d** OFV (blue line) of a narrow linewidth fiber laser (NKT Photonics Basik E15) being sinusoidally modulated at 100 kHz and the corresponding scan rate (red line) obtained by taking the derivative of the OFV data showing that a frequency variation speed of over ±25 THz/s can be clearly measured. The data shown in **d** has been digitally filtered with a low pass filter (2nd order Butterworth with a cutoff frequency of 200 kHz) to remove excessive electronic noises. OFV: optical frequency variation.

    Fig. 5a shows the measured optical frequency of a rapidly tunable external cavity laser (Newport TLB-8800-HSH-CL) scanned at the highest rate of 20000 nm/s (2500 THz/s) with

data acquired at a rate of 62.5 MS/s with two digital oscilloscopes (Rohde & Schwarz RTB2004, 10 bit resolution, 4 channels each), while the inset shows the zoom-in view of the data in the black box, revealing the details of the stairway-like frequency increments as a function of time. It is evident that not only is our PIC OFD capable of measuring such high-speed frequency sweeps, but also the details of the OFV dynamics, which possibly reflects the effect of the step-motor based laser tuning mechanism. f-clock (or k-clock) can be generated from the OFV data in a by digitally outputting pulses with a frequency increment of user's choice, as shown in Fig.5b, in which a frequency increment of 0.5 GHz is chosen to faithfully represent the local OFV rate. Fig. 5c is the wavelength scan rate derived from the data in Fig. 5a, which shows that it deviates from the setting rate of 20000 nm/s and fluctuates at different times during the scan. Such a deviation is identified as the laser tuning imperfections, not the measurement artifacts, which can be characterized in detail by our PIC OFD.

To further validate the high-speed frequency measurement capability of our PIC OFD, we sinusoidally modulate the frequency of a narrow linewidth laser (NKT Photonics Basik E15, 1550.12 nm, 16.2 dBm output power) with a modulation depth setting of 100% at a rate of 100 kHz using the laser's built-in frequency modulation function and measure the corresponding OFV with our PIC OFD. The results are shown by the blue curve in Fig. 5d, which can be further used to obtain the rate of the OFV by taking the derivative of the data, as shown by the red curve in Fig. 5d. It is evident that our PIC OFD can provide detailed information on the rapidly varying optical frequency and its variation rate.

### *High-resolution Optical Frequency Variation measurements*

To demonstrate the fine detail measurement capability of our PIC OFD, a tunable laser fine-tuned with the minimum step size of 1 pm is measured, with the results shown in Fig. 6a. It is evident that our PIC OFD is more than sufficient to resolve the 1 pm steps. More importantly, it can even detect the transient dynamics when the wavelength steps from one wavelength to another, with the overshoots and settling down processes clearly displayed. In addition, the deviations of wavelength step sizes from the setting step of 1 pm can be clearly observed, which fluctuate between 0.74 pm and 1.76 pm due to the imperfections of the tunable laser, as shown by the measured step sizes in Fig. 6a.

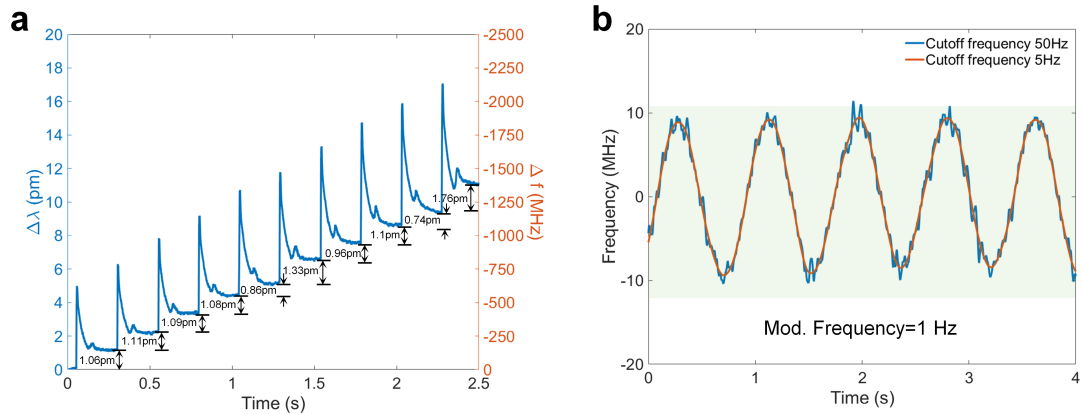

**Fig. 6. High resolution OFV measurement capability demonstration. a** The measured wavelength (frequency) as a laser (Yenista TUNICS T100S-HP) is step-tuned at 1 pm per step. **b** The measured optical frequency variation when the frequency of a fiber laser (NKT Photonics BASIK E15) is modulated with a sinusoidal waveform at 1 Hz, with the data digitally filtered by a low pass filter (3rd order Butterworth) having a cutoff frequency of 50 Hz (blue) and 5 Hz (red), respectively.

To demonstrate the even finer OFV measurement capability of our PIC OFD, the same NKT fiber laser used for Fig. 5d with a much finer frequency tuning range is used to generate the required OFV for the PIC OFD to measure, with the result shown in Fig. 6b. The laser is set at the sinusoidal frequency modulation mode with a modulation frequency of 1 Hz and a modulation depth of 10%. The same DAQ card used in Fig. 3 is also used here. As can be seen in Fig. 6b, OFVs with an amplitude of 10 MHz (0.08 pm) can be clearly measured at a measurement bandwidth of 50 Hz and a frequency resolution on the order of 2 MHz (0.016 pm) can be discerned at a detection bandwidth of 5 Hz. The measurement of higher speed OFVs requires higher bandwidth, which may compromise the frequency measurement resolution due to increased noise level, as will be discussed in more detail in the last section.

### PIC OFD for FBG interrogation

The high resolution and high-speed OFV detection capability of the PIC OFD can be utilized to interrogate fiber Bragg grating (FBG) sensors[39,41,42], as will be demonstrated in this section. Fig. 7a shows the experiment setup, which includes an ASE broadband light source, an optical circulator, and an FBG mounted on a self-made cantilever beam with adhesives. The reflected light from the FBG is amplified by an erbium doped fiber amplifier (EDFA) before entering the PIC OFD. The FBG has reflection bandwidth of 0.26 nm (corresponding to a coherence length of 9.2 mm), with a strain sensitivity of 1.048 pm/με, calibrated with a strain sensor.

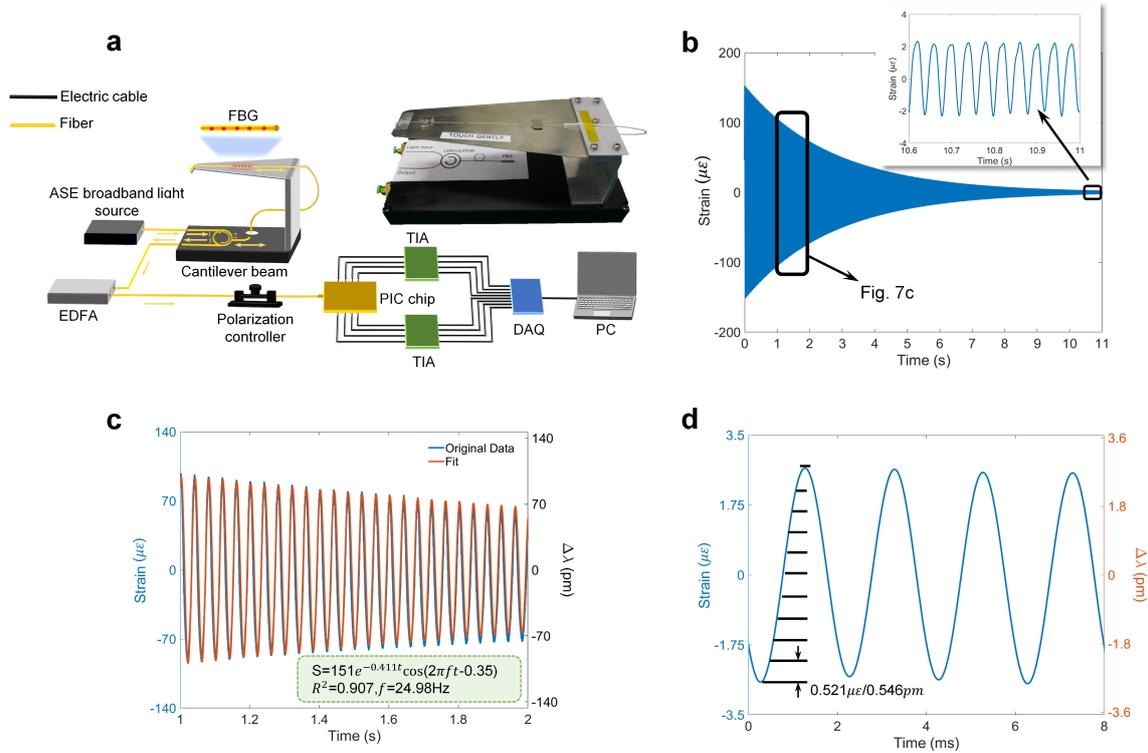

**Fig. 7. Demonstration of FBG interrogation with PIC OFD. a** Experiment setup with a (FBG: Fiber Bragg Grating) placed onto a self-made cantilever beam (inset). **b** The measured dynamic strain showing the damped oscillation of the cantilever beam excited by tapping it with a fingertip, with the data filtered by a 3rd order Butterworth filter having a cutoff frequency of 100 Hz. Inset: expanded view of the data in the black box around 10 seconds. **c** Details of the damped oscillation and the corresponding curve fit between 1 and 2 seconds. **d** Measured dynamic strain induced by the acoustic wave from a speaker underneath the cantilever beam, with the data filtered by a 3rd order Butterworth filter having a cutoff frequency of 1 kHz.

In operation, the center wavelength of the reflected light from the FBG changes linearly with the strain applied to the FBG[43] via the cantilever beam, which can be detected by the PIC OFD with the same low cost DAQ card (60 kS/s and 16-bit resolution) used in Fig. 3. The strain can be obtained from the amount of center wavelength shifts using the strain sensitivity of 1.048 pm/με. Fig. 7b shows the dynamic strain variation with time when the cantilever beam is tapped by fingertip. The damped oscillation of the cantilever beam can be clearly seen, which can be fitted to a typical damped oscillation function with a time constant of 0.411/s and a resonant frequency of 24.98 Hz, as shown in Fig. 7c. The oscillation lasted more than 12 seconds before it was too small to be detected, with the zoom-in strain oscillation data from 10.6 to 11 seconds displayed in the inset of Fig. 7b. Fig. 7c is the expanded view of the data in Fig. 7b marked with the black box, showing the details of the damped strain oscillation.

To further demonstrate the measurement sensitivity and speed of our PIC OFD, a small speakerphone (Yealink CP700 with a frequency range of 150 Hz~8 kHz) was placed underneath the cantilever beam, which was driven by a 500 Hz sinusoidal signal from a smart phone. The detected dynamic strain is shown in Fig. 7d, which has a strain variation amplitude of 2.6 με and a frequency of 500 Hz, consistent with the driving signal. A strain resolution on the order of 0.1~0.2 με at 500 Hz can be discerned from Fig. 7d, which is an order of magnitude finer than the best commercial FBG interrogators, as will be shown discussed in the Summary and Discussion section below.

## Discussion

In summary, we report a photonic integrated sine-cosine optical frequency detector, consisting of a pair of unbalanced I-Q interferometers: a main interferometer having a small FSR (or large OPD) and an assistive interferometer with large FSR (or small OPD). The main interferometer is for achieving high resolution, while the assistive interferometer is for estimating the absolute frequency or wavelength, which in turn is used to estimate the absolute wavelength and determine the wavelength dependent parameters for the main interferometer.

The device is realized on the TFLN platform with the large OPD of only 10 mm and a total size of 5.5 mm x 2.7 mm, which is shown to be able to measure OFVs down to 2 MHz (0.016 pm) at 5 Hz, limited by the excessive electronic circuit noise in our measurement system. Such a fine resolution enabled us to see and quantify minor frequency scan imperfections, such as ripples, overshoots, and nonlinearities, which otherwise could not be detected or even noticed with other measurement methods. With a data acquisition rate of 62.5 MS/s, we demonstrated the accurate measurement of frequency (wavelength) variation speed up to 2500 THz/s (20000 nm/s), limited by the achievable laser wavelength scan rate. The minor frequency scan details at such high speed, such as the scanning steps and scanning rate irregularities, can also be clearly detected and quantified. All the frequency (wavelength) scanning characteristics of tunable lasers, including the direction, periodicity, scanning waveforms, scanning rates, nonlinearity, ripples, and repeatability can be accurately determined in the wavelength range of 160 nm from 1480 nm to 1640 nm. The superb performances are ensured by the robust mathematics and associated calibration procedures we developed, which fully accounts for the wavelength dependent imperfections of the device resulting from the design and fabrication tolerances and makes it feasible for being widely adopted in real world applications.

As discussed in[19,44], the OFV resolution of our PIC OFD is limited by either the resolution of the DAQ card or the detector noise (including the thermal, shock, and electronic noises), whichever is larger, which can be written as[19]:

$$\delta f_{DAQ} = \frac{FSR_u/2}{2^m} = \frac{1}{\tau_u 2^{m+1}} \tag{1}$$

$$\delta f_n = \frac{FSR_u/2}{V_{pp}/\delta V_n} = \frac{1}{2\tau_u SNR_V} = \frac{1}{2\tau_u \sqrt{SNR_P}} = \frac{1}{2\tau_u} \sqrt{\frac{\rho_n B}{P_s}} \tag{2}$$

where $\delta f_{DAQ}$ and $\delta f_n$ are the OFV resolution limited by the DAQ card and the system noise, respectively, m is the effective bit number of the DAQ card, $V_{pp}$ is the peak-peak voltage of the circuit output, $\delta V_n$ is the voltage noise, B is the bandwidth of the detector circuitry, including the PD and TIA, $\rho_n$ is the noise power spectral density, $P_s$ is the signal power, and $SNR_V$ and $SNR_P$ are the voltage and power signal to noise ratios (SNR), respectively.

On the other hand, the measurable OFV speed or chirp rate $\sigma_{max}$ is limited by the sampling rate $R_s$ of the DAQ card used, which can be written as[19]:

$$\sigma_{max} = \frac{R_s}{2\tau_u} = \frac{B}{\tau_u} = B \cdot FSR_u \tag{3}$$

where $R_s = 2B$ is taken. Combining the two equations yields the OFV detection resolution $\delta f_n$ corresponding to the required OFV measurement speed $\sigma_{max}$:

$$\delta f_n = \frac{1}{2} \sqrt{\frac{(\rho_n/\tau_u)\sigma_{max}}{P_s}} = \frac{1}{2} \sqrt{\frac{(\rho_n \cdot FSR_u)\sigma_{max}}{P_s}} = \frac{1}{2} \sqrt{\frac{\sigma_{max}}{SNR'_P}} \tag{4}$$

where $SNR'_P = \frac{P_s}{\rho_n \cdot FSR_u}$ is the power SNR in the bandwidth of the FSR of the upper (main)

interferometer. Clearly, the OFV resolution due to the noise is limited by the required OFV detection speed $\sigma_{max}$ and the SNR in the bandwidth of the FSR of the main unbalanced I-Q interferometer.

**Table 1.** Limits of OFV measurement speeds and resolutions at different detection bandwidths.

| | OFV Resolution $\delta f$ | | | |
|---|---|---|---|---|
| Detection BW (Hz) | $\sigma_{max}$ THz/s (nm/s) | 16-bit DAQ ($\delta f_{DAQ}$) | Fundamental noises $\delta f_n$ ($\delta V_n$) | Commercial electronics $\delta f_n'$ ($\delta V_n'$) |
| 500 M | $1.5 \times 10^7$ ($1.2 \times 10^8$) | | 1.09 MHz (0.29 mV) | 6.49 MHz (1.73 mV) |
| 100 M | $3 \times 10^6$ ($2.4 \times 10^7$) | 0.458 MHz | 0.49 MHz (0.13 mV) | 3.41 MHz (0.91 mV) |
| 10 M | $3 \times 10^5$ ($2.4 \times 10^6$) | | 153 kHz (40.8 μV) | 1.13 MHz (0.3 mV) |
| 1 M | $3 \times 10^4$ ($2.4 \times 10^5$) | | 48.4 kHz (12.9 μV) | 0.37 MHz (0.1 mV) |

Table 1 lists the theoretically achievable OFV detection speeds $\sigma_{max}$, the resolutions $\delta f_{DAQ}$ limited by the DAQ card resolution, the resolution $\delta f_n$ limited by fundamental noises, and the resolution $\delta f_n'$ limited by the noises of a commercial available electronics (PD, TIA and DAQ card) at different detection bandwidths, for our PIC OFD having an FSR of 30 GHz. In the calculation, the TIA is assumed to have an output voltage range of 4.0 volts and the DAQ card is assumed to have an effective resolution of 15-bits capable of digitizing voltages in a range of 4 volts. As can be seen from Table 1, for a perfect electronic circuit with fundamental noise levels ($\delta V_n$ in parentheses), the OFV resolution is 458 kHz when detection bandwidths less than 100 MHz, determined by the resolution of the DAQ card. Above 100 MHz detection bandwidth, the OFV resolution is determined by the fundamental noise, at 490 kHz and 1.09 MHz, corresponding to detection bandwidths of 100 MHz and 500 MHz, respectively. However, in practice it is difficult if not impossible for the electronic noises to be lowered to the fundamental noise levels. The last column of Table 1 lists the OFV detection resolution $\delta f_n'$ limited by the noise levels ($\delta V_n'$ in parentheses) of a commercial photodetector with TIA (Thorlabs' PBD470C) measured with an oscilloscope (Rohde & Schwarz RTB2004). At a detection bandwidth of 1MHz or less, the OFV detection resolution is 458 kHz, determined by the DAQ card with an effective resolution of 15-bit. At a detection bandwidth of more than 1MHz, the OFV detection resolution is determined by the electronics noise levels $\delta V_n'$.

**Table 2.** Performance comparison of different FBG interrogators.

| Brand and Model | Wavelength resolution | Strain resolution | Temperature resolution | Wavelength range(nm) | Sampling rate | Interrogator mechanism |
|---|---|---|---|---|---|---|
| Smart Fibres SmartScan | 5 pm | 4.2 $\mu\varepsilon$ | 0.45 ℃ | 1528~1568 | 2.5 kHz | Tuneable laser source |
| Geokon Instrument BGK-FBG8600L | 3 pm | 2.5 $\mu\varepsilon$ | 0.27 ℃ | 1525~1565 | 2 Hz | Tunable fiber lasers and spectral analysis |
| Luna si255-EV-08-1460-1620-0010-DP | 1 pm | 0.83 $\mu\varepsilon$ | 0.089 ℃ | 1460~1620 | 10 Hz | Fiber Fabry-Perot filter and wavelength reference |
| FBGS FBG-Scan 90X | 0.3 pm | 0.25 $\mu\varepsilon$ | 0.027 ℃ | 1510~1590 | 1 kHz | Spectrometer |
| PIC AWG interrogator[4] | 2 pm | -- | -- | 1530~1625 | -- | Arrayed Waveguide Grating (AWG) |
| PIC active modulation interrogator[41] | 0.0873 pm | -- | -- | -- | 7.21 kHz | Active and passive phase sensitive detections |
| PIC sine-cosine OFD (at 5Hz BW) | 0.016 pm | 0.013 $\mu\varepsilon^*$ | 0.0015 ℃* | 1480~1640 | 60 kHz Demonstrated, higher feasible | Sine-cosine encoder |
| PIC sine-cosine OFD (16-bit ADC limit)** | 0.0037 pm | 0.003 $\mu\varepsilon$ | 0.34×10⁻³ ℃ | | | |

\* Calculated with the wavelength resolution of 0.016 pm (circuit noise limited results).

\*\* Wavelength resolution calculated using Eq. (1), assuming a 16-bit DAQ with 15 bits effective resolution.

As an application example, we also demonstrated that our sine-cosine PIC OFD can be used as a high speed and high resolution FBG interrogator, with a strain measurement resolution of 0.1~0.2 µε at 500 Hz. Corresponding to the wavelength resolution of 0.016 pm at 5Hz measurement bandwidth, the calculated strain and temperature resolution are 0.013 µε and $0.0015$ ℃, respectively, which is about 20 times more sensitive than those of the best commercial interrogators, as shown in Table 2, and can be used to measure the dynamic strain variations induced by vibrations and acoustics with high sensitivity. Even better strain and temperature resolution can be achieved if the OFV resolution of our PIC OFD can be improved by reducing the circuit noise to the level achievable with commercial electronics. As indicated in Table 1, for a detection bandwidth up to 1 MHz (corresponding to an ultra-high OFV detection speed $\sigma_{max}$ of 3×10⁴ THz/s), the OFV resolution is limited by the resolution of the 16-bit DAQ card (with 15-bit effective resolution) to be 458 kHz (0.0037 pm), corresponding to a strain and temperature resolutions of 0.003 µε and 0.34×10⁻³ ℃, respectively, as shown in the last row of Table 2, which are over 80 times more sensitive than those of the best commercial FBG interrogators.

The present PIC OFD suffers from high insertion loss and high wavelength dependent loss, mostly due to the grating coupler used for attaching the input fiber, which can be improved significantly in our next tape out using edge coupling with a properly designed spot size converter. In addition to the noises in the TIA, the back reflection from the photodectors (PD) may also affect the performance of our OFD. In our experiments, Indium Gallium Arsenide (InGaAs) PD chips are butt-coupled and bonded to the waveguide output with adhesives, which may cause back reflection induced interferometric ripples to degrade the OFD performance, as well as add packaging cost. Recent progress shows that Indium phosphide (InP) and InGaAs photodetectors can be heterogeneously integrated on to TFLN PIC[45] for applications in the wavelength range of 1200 nm to 1650 nm. In addition, black phosphorus photodetector can also be hybridly integrated on TFLN waveguide[46] around 1550 nm. For applications in the visible and near infrared wavelength range, silicon photodetectors can be integrated in the TFLN PIC[47]. Alternatively, if silicon on insulator (SOI) or silicon nitride (SiN) platforms is used for fabricating the OFD PIC, the Germanium (Ge) PDs or Indium Selenide (InSe) PDs can be monolithically integrated in the PIC[48-50]. The integration of such PDs in the sine-cosine OPD PIC may avoid the back reflection issue mentioned above for performance improvement while reducing packaging cost.

Furthermore, the long leads connecting the PD outputs to the TIAs picked up a large amount noises from the power supplies in our experiment, causing large noise spikes at 50 Hz and its harmonics, which greatly degraded the OFV measurement resolution. It is feasible to directly attach TIA chips onto PD outputs in the OFD PIC using hybrid integration to further improve measurement resolution and accuracy.

The present sine-cosine OFD is designed for S to L band operation, covering a broad bandwidth of 160 nm from 1480 nm to 1640 nm. The wavelength range of the design can be readily extended to O band around 1310 nm and even into visible range because TFLN is transparent from 350 nm to 5500 nm[51], such that different devices can cover different spectral bands. It is also possible to extend the wavelength range of a single device, however, the bandwidth range of all the components in the OFD, such as the waveguides, the couplers, and the MMI, may also need to be modified to allow operation in a wider wavelength range around

240 nm[38].

Much finer OFV resolution can be achieved with a smaller FSR or longer OPD in the main I-Q interferometer if required, possibly on the SiN platform with lower waveguide transmission loss, because the resolution scales linearly with the FSR (or the OPD) of the I-Q interferometer, as indicated by Eq. (1). For example, the OFV resolution can be improved 100 times to 20 kHz by reducing the FSR of the main I-Q UMZI by 100 times or increasing its OPD to 1 meter (by 100 times), assuming the same relatively noisy electronics in the present experiment will be used. Such a high OFV resolution is entirely feasible considering that SiN waveguide with a low loss on the order of 1 dB/m can be readily achieved[52]. The OFV resolution of this OFD can be further improved to 4.6 kHz by using lower noise electronics, as indicated in Table 1, assuming that it is limited by the DAQ of 15-bit effective resolution. If a DAQ of 18-bit effective resolution is used, an OFV resolution of 0.58 kHz can be achieved.

Our work on the PIC OFD and the associated robust demodulation algorithm point to a fruitful direction for the on-chip OFV measurement with high resolution, high speed and wide wavelength range, which is attractive for on-chip laser frequency control or stabilization, photonic computing, FBG interrogation, in addition to the k-clock generation on the PIC chips for OFDR, OCT, FMCW Lidar, and TDLS applications, as depicted in Figs. 8a-8d.

For example, k-clock generation is generally used in OFDR, FMCW Lidar, and OCT systems for improving system spatial resolution by compensating for the laser scanning nonlinearity[2, 53-57]. Comparing with the k-clock generation using a conventional UMZI, the OPD can be reduced by $1/2^m$, as indicated by Eq. (1), making it possible to integrate the k-clock on the same sensing chip, as shown in Fig. 8a. As an added benefit, the detected signal speed corresponding to a given OFV is also reduced by $1/2^m$, which can significantly lower the bandwidth and hence the noise of the electronics. In OFDR and FMCW Lidar systems, OPDs on the order of few hundred meters to tens of kilometers are generally required[2] using a conventional UMZI for k-clock generation. For example, for an OFDR or FMCW Lidar system, a commonly required OPD of 300 m can be reduced to merely 9.16 mm with our sine-cosine OFD design (assuming $m = 15$), which is sufficiently small to be integrated on a PIC. Even an OPD requirement of 15 km for k-clock generation with a conventional design[58] can be effectively reduced to 45.8 mm, which can be conveniently made on a low loss TFLN or SiN platform.

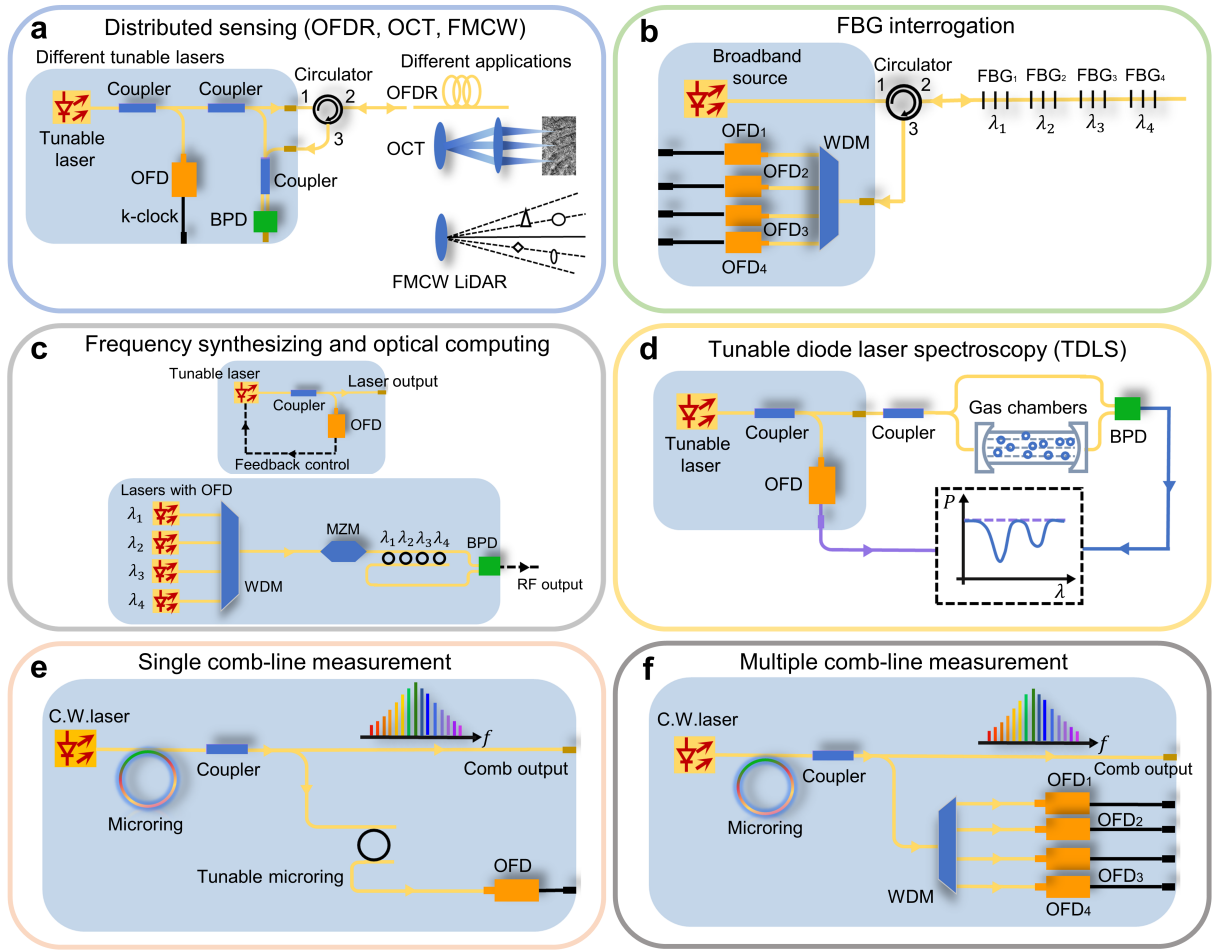

**Fig. 8. Application illustrations of the on-chip OFD**. **a** On-chip k-clock generation for different distributed optical sensing systems (OFD: optical frequency detector, OFDR: optical frequency domain reflectometer, OCT: optical coherence tomography, FMCW LiDAR: frequency modulated continuous wave LiDAR , BPD: Balanced photodetector). **b** Chip-sized multi-channel fiber Bragg grating (FBG) interrogator (WDM: wavelength division multiplexing). **c** On-chip frequency monitoring and control of integrated lasers for frequency synthesizing and MRR based optical computing[8] (MZM: Mach-Zehnder modulator). **d** For rapidly obtaining the laser wavelengths in TDLS systems. **e** The instantaneous frequency measurement of a single comb-line in a microcomb. **f** Simultaneous frequency measurement of multiple comb lines in a microcomb.

In addition to the applications depicted in Figs. 8a to 8d, the sine-cosine OFD may also be used to measure or monitor the frequencies of an optical frequency comb by integrating a thermally tunable MRR filer[59] in front of the OFD to select a particular comb line to measure its instantaneous frequency, as illustrated in Fig. 8e. Alternatively, a wavelength division multiplexer (WDM) implemented with arrayed waveguide grating (AWG)[60] or cascaded Mach-Zehnder interferometers (MZI)[61] can also be integrated in front of multiple OFDs to separate different comb lines into different OFDs to simultaneously measure the instantaneous frequencies these comb lines, as shown in Fig. 8f. The OFD or OFDs may also be integrated on the same chip with a microcomb for comb line monitoring and control.

## Methods

### *Device design and fabrication*

The proposed device was fabricated using an x-cut LNOI wafer, which consists of a 360-nm-thick X-cut LN layer and a 4.7-μm-thick buried oxide layer. The device includes an edge coupler, three 3-dB 2×2 bend-directional couplers (BDCs), two 90° hybrids, and eight photodetectors to form the upper and lower interferometers. The OPDs for the two interferometers are designed to be 10 mm and 14.43 μm, corresponding to a FSR of 30 GHz and 20.79 THz (169.3 nm), respectively. It is packaged in a butterfly case with a thermoelectric cooler (TEC) and the temperature is set to 30 °C during normal operation to avoid OPD changes due to temperature changes.

The fabrication process is simple and straightforward, which will be described in the Supplementary Information Section 4 of the paper. Such a simple fabrication process ensures that the device can be mass produced with high yield at low cost.

## Data Availability

Source data presented in this paper, and main data supporting the findings of this study are available in the Supplementary Information files.

## Acknowledgements

This work was supported in part by the National Natural Science Foundation of China (62405382 to Z.L., 62293523 to X.C); Natural Science Foundation of Hebei Province (F2024201002 to X.S.Y); Interdisciplinary Research Program of Natural Science of Hebei University (DXK202204 to X.S.Y); Advanced Talents Program of Hebei University (521000981006 to X.S.Y); Internal R & D funding of NuVision Photonics, Inc.

## Author Contributions Statement

X.S.Y. and X.C. initiated and supervised the project. X.S.Y. proposed and designed the photonic integrated circuit (PIC) chip for measuring rapid optical frequency variations (OFV) and drafted the initial manuscript. Y.L. and X.M. conducted all measurements and data processing. Z.L. designed the layout of the PIC chip. Y.Z., W.K., H.T., and X.W. completed the fabrication of the PIC chip and participated in the discussion of the results. X.S.Y., Y.L., X.M., Z.L., and X.C. discussed the results and edited the manuscript together.

## Competing Interests Statement

The authors declare no competing interests.

# Supplementary Information for "On-chip real-time detection of optical frequency variations with ultrahigh resolution using the sine-cosine encoder approach"

X. Steve Yao[1,2,*], Yulong Yang[1], Xiaosong Ma[1], Zhongjin Lin[3], Yuntao Zhu[3], Wei Ke[4], Heyun Tan[3], Xichen Wang[5], and Xinlun Cai[3,*]

[1]Photonics Information Innovation Center and Hebei Provincial Center for Optical Sensing Innovations, College of Physics Science and Technology, Hebei University, Baoding 071002, China

[2]NuVision Photonics, Inc, Las Vegas, NV 89109, USA

[3]State Key Laboratory of Optoelectronic Materials and Technologies, School of Electronics and Information Technology, Sun Yat-sen University, Guangzhou 510275, China

[4]Liobate Technologies, Guangzhou, China

[5]NeoPIC Technologies, Suzhou, China

*Corresponding authors: X. Steve Yao syao@ieee.org and Xinlun Cai caixlun5@mail.sysu.edu.cn

# 1. More detailed device description

In our implementation, each 90° hybrid is made with a 2×4 multimode interference (MMI) coupler[1], although other types of 90° hybrid can also be implemented[2-6]. The eight output ports of the two MMI couplers are attached with eight photodetectors (PD) labeled PD1 through PD8 for detecting the corresponding optical powers $P_i(t)$ ($i = 1,2 \ldots 8$) and then converting them to eight voltages via eight transimpedance amplifiers (TIA). The four outputs in each interferometer can be grouped into two pairs. Ideally, the two interference signals in each pair are 180° out of phase, while the phase difference between the two pairs is 90°. Therefore, one pair of outputs is often called cosine channels (in-phase channels or I-channels), while the other pair is called sine channels (quadrature channels or Q-channels). That is why the scheme is called sine-cosine scheme, or alternatively, I-Q scheme. For simplicity, we may call the interferometers in Fig. 1 unbalanced I-Q interferometers.

In practice, the phase difference between the two signals in each pair is not perfectly 180°, and between the two pairs is not perfectly 90°. Therefore, the four output photovoltages $V_{uj}$ ($j = 1,2,3,4$) from the upper (main) interferometer can be expressed in general as:

$$V_{u1}(t) = A_{u1} + B_{u1}(\lambda)\{1 + m_{u1}\cos\left[\Delta\varphi_u(t) - \frac{\alpha_u(\lambda)}{2}\right]\} \tag{S1}$$

$$V_{u2}(t) = A_{u2} + B_{u2}(\lambda)\{1 - m_{u2}\cos\left[\Delta\varphi_u(t) + \frac{\alpha_u(\lambda)}{2}\right]\} \tag{S2}$$

$$V_{u3}(t) = A_{u3} + B_{u3}(\lambda)\{1 + m_{u3}\sin\left[\Delta\varphi_u(t) + \gamma_u(\lambda) - \frac{\beta_u(\lambda)}{2}\right]\} \tag{S3}$$

$$V_{u4}(t) = A_{u4} + B_{u4}(\lambda)\{1 - m_{u4}\sin\left[\Delta\varphi_u(t) + \gamma_u(\lambda) + \frac{\beta_u(\lambda)}{2}\right]\} \tag{S4}$$

where the subscript "u" stands for the upper interferometer, $A_{uj}$ are the voltage bias due to the imperfection of the electronic circuit for each channel, $B_{uj}$ are the amplitudes of the sine and cosine functions relating to the received optical power in each channel which are weakly wavelength dependent, $m_{uj}$ are the modulation depth of each channel, and $\alpha_u$, $\beta_u$, $\gamma_u$ are the phase imperfections of the 90° hybrid, which are also weakly wavelength dependent in general and are zero if the 90° hybrid is perfect. Finally, $\Delta\varphi_u(t)$ is the phase induced by the OFV $\Delta f(t)$ and can be expressed as:

$$\Delta\varphi_u(t) = 2\pi\Delta f(t)\tau_u = \frac{2\pi\Delta f(t)}{FSR_u} = \frac{2\pi\Delta f(t)}{c/\Delta L_u} \tag{S5}$$

In Eq. (S5), c is the speed of light, $\tau_u$ is the time delay between the two arms of the upper

I-Q interferometer, $\Delta L_u$ is the corresponding OPD, and $FSR_u$ is the corresponding free spectral range, which are related by the following expression:

$$FSR_u = 1/\tau_u = c/\Delta L_u \tag{S6}$$

Note that $FSR_u$ (or $\tau_u$) may vary with wavelength due to dispersion. In practice, $FSR_u$ as a function of $\lambda$ can be accurately obtained with a tunable laser by measuring the periodicities of the sinusoidal interference signals of Eqs. (S1) to (S4) when the wavelength of the laser is tuned, as shown in Supplementary Information Section 2.

Similar expressions can also be obtained for the lower I-Q interferometer by simply replacing the subscript "$u$" with "$l$" in Eqs. (S1) to (S5), where "$l$" stands for the lower interferometer.

If the 90° hybrid is perfect, $\alpha_u = \beta_u = \gamma_u = 0$. However, in practice, these parameters are non-zero and wavelength dependent. In such non-ideal situations, a calibration procedure must be performed first to obtain these non-zero parameters, including $B_{uj}$ and $m_{uj}$, as a function of wavelength, which can be tabulated in a lookup table containing ($B_{uj}$, $m_{uj}$, $\alpha_u$, $\beta_u$, $\gamma_u$) vs. $\lambda$, in addition to $FSR_u$ vs. $\lambda$, as will be described in detail in Supplementary Information Section 2 below. Once these parameters are determined, the phase increment $\Delta\varphi_u(t)$ and the corresponding frequency increment $\Delta f(t)$ can be obtained from Eqs. (S1) to (S4) as

$$\cos[\Delta\varphi_u(t)] = \frac{v_{u1}(t)-v_{u2}(t)}{2\cos\frac{\alpha_u}{2}} \tag{S7}$$

$$\sin[\Delta\varphi_u(t)+\gamma_u] = \frac{v_{u3}(t)-v_{u4}(t)}{2\cos\frac{\beta_u}{2}} \tag{S8}$$

$$\Delta f(t) = \frac{1}{2\pi\tau_u}\Delta\varphi_u(t) = \frac{1}{2\pi\tau_u}tan^{-1}\left\{\frac{[v_{u3}(t)-v_{u4}(t)]\cos\frac{\alpha_u}{2}-[v_{u1}(t)-v_{u2}(t)]\cos\frac{\beta_u}{2}\sin\gamma_u}{[v_{u1}(t)-v_{u2}(t)]\cos\frac{\beta_u}{2}\cos\gamma_u}\right\} \tag{S9}$$

where $v_{uj} = [V_{uj}(t) - A_{uj} - B_{uj}]/(m_{uj}B_{uj})$ ($j = 1,2,3,4$) are normalized voltages with the circuit bias $A_{uj}$ and the interferometer DC term $B_{uj}$ subtracted. Like the sine-cosine motor encoder, the direction of the frequency variation can also be determined[19], which allows the unwrapping of phase changes over $2\pi$. Note that because all the device imperfections have been considered in Eq. (S9), the formula is extremely robust against design and fabrication tolerances, which ensures the feasibility for mass deployment of the scheme in real world applications. The obtained $\Delta\varphi_u(t)$ and the corresponding $\Delta f(t)$ is depicted in Fig. 1d.

Note that the upper interferometer by itself can only measure the incremental frequency, not the absolute frequency because of the multiple periodic cycles of the output signal corresponding to the OFVs in the measurement range. To estimate the absolute frequency or wavelength, the lower interferometer can be used, because the large FSR assures a single value output in the intended measurement range, as shown in Fig. 1. Like Eqs. (S7) to (S9), the phase $\Delta\varphi_l(\lambda)$ in the lower unbalanced I-Q interferometer corresponding to a wavelength $\lambda$ can be written as:

$$\Delta\varphi_l(\lambda) = tan^{-1}\left\{\frac{[v_{l3}(\lambda)-v_{l4}(\lambda)]\cos\frac{\alpha_l}{2}-[v_{l1}(\lambda)-v_{l2}(\lambda)]\cos\frac{\beta_l}{2}\sin\gamma_l}{[v_{l1}(\lambda)-v_{l2}(\lambda)]\cos\frac{\beta_l}{2}\cos\gamma_l}\right\} \tag{10}$$

where $v_{lj} = [V_{lj}(t) - A_{lj} - B_{lj}]/(m_{lj}B_{lj})$ $(j = 1,2,3,4)$ are normalized voltages with the circuit bias $A_{lj}$ and the interferometer DC term $B_{lj}$ subtracted, $\alpha_l, \beta_l, \gamma_l$ are the phase imperfections of the 90° hybrid in the lower interferometer. One may use a desk-top tunable laser with precisely known wavelengths to calibrate the output of the lower interferometer by relating $\Delta\varphi_l(\lambda)$ obtained using Eq. (S10) with the wavelength of the tunable laser, and then store the results of $\Delta\varphi_l$ vs. $\lambda$ in a lookup table. This way, at the start of optical frequency or wavelength measurement, the absolute frequency $f_0$ or wavelength $\lambda_0$ can be determined once $\Delta\varphi_l$ is calculated from the four outputs of the lower interferometer. Combining with Eq. (S9), one gets:

$$f(t) = f_0 + \Delta f(t) \tag{S11}$$

Perhaps more importantly, the lookup table of $\Delta\varphi_l$ vs. $\lambda$ can be used for determining $\lambda$ for the selection of the wavelength dependent parameters in the lookup table of $(B_{uj}, m_{uj}, \alpha_u, \beta_u, \gamma_u, FSR_u)$ vs. $\lambda$ for calculating $\Delta f(t)$ with Eq. (S9), which will be discussed in more details in Supplementary Information Section 3.

Note that corresponding to an OFV, the signal variation speeds from the four output channels in the upper interferometer are $\eta$ times faster than those from the lower interferometer, where m is their FSR ratio ($\eta = \frac{FSR_l}{FSR_u}$). Therefore, the TIAs and the DAQ card for the lower interferometer can be $\eta$ times slower than those for the upper interferometer for significantly reduced cost, although in most of our experiments, all the channels of the DAQ card have the same speed of 60 kS/s.

## 2. Upper (main) I-Q interferometer calibration

As described in Device Calibration subsection, the fabricated device is not perfect, with many parameters deviating from ideal values and wavelength dependent, which must be obtained by calibration at different wavelengths. The setup for calibrating and evaluating the sine-cosine PIC OFD is shown in Fig. 2. The circuit bias $A_{uj}$ in Eqs. (S1) to (S4) can be obtained by simply turning the laser off. By scanning the laser frequency in the vicinity of $\lambda$ with a range larger than $FSR_u$ (the FSR of the upper or main I-Q interferometer), four interference signals $V'_{uj}(t)$ ($j = 1,2,3,4$) of Eqs. (S1) to (S4) can be obtained, as shown in Fig. S1a for $\lambda = 1519$ nm, where $V'_{uj}(t) = V_{uj}(t) - A_{uj}$. By plotting the Lissajous figures of $V'_{u1}(t)$ vs. $V'_{u3}(t)$, $V'_{u2}(t)$ vs. $V'_{u3}(t)$, and $V'_{u1}(t)$ vs. $V'_{u4}(t)$, and finally performing the elliptical fits of these three Lissajous figures, all the wavelength dependent parameters ($A_{uj}$, $B_{uj}$, $m_{uj}B_{uj}$, $\alpha_u$, $\beta_u$, $\gamma_u$) at $\lambda = 1519$ nm can be obtained. Fig. S1b shows an example of plotting $V'_{u1}(t)$ vs. $V'_{u3}(t)$ and performing elliptical fit, with which $B_{u1} = 1.0958$ V, $B_{u3} = 0.9177$ V, $m_{u1} = 0.976$, $m_{u3} = 0.95$, and $\gamma_u - \frac{\beta_u}{2} + \frac{\alpha_u}{2} = 2.659°$ are obtained. By completing the remaining two elliptical fits, $\alpha_u$, $\beta_u$, $\gamma_u$ and $B_{u3}$ can be obtained.

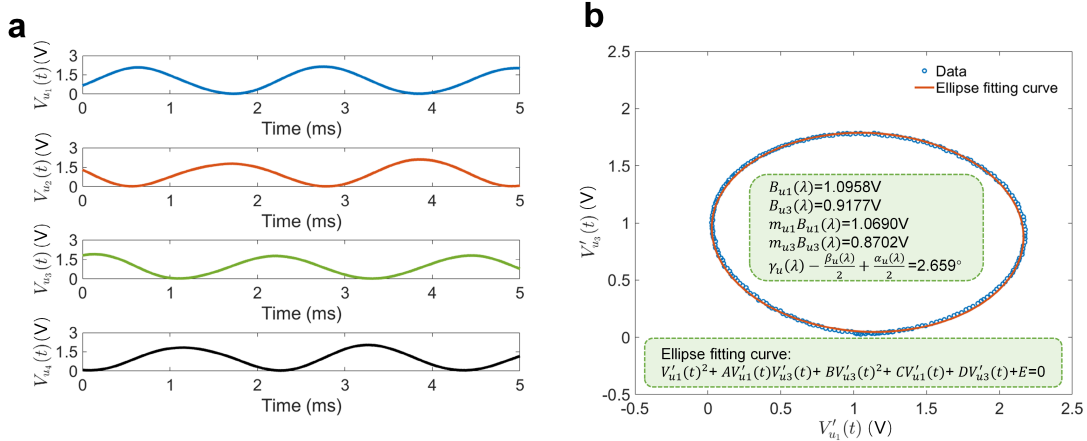

**Fig. S1. Lissajous fitting for determining demodulation parameters. a** The four output interference signals from the main I-Q interferometer when the tunable laser is tuned in the vicinity of 1519 nm. **b** The Lissajous figure of $V'_{u1}(t)$ vs. $V'_{u3}(t)$ and the elliptical fit.

By stepping the wavelength $\lambda$ every 1nm, a lookup table of ($A_{uj}$, $B_{uj}$, $m_{uj}B_{uj}$, $\alpha_u$, $\beta_u$, $\gamma_u$) vs. $\lambda$ can be obtained, as shown in Fig. S2, which can be used to calculate the OFVs of unknown light sources using Eq. (S9). Fig. S2a shows the obtained wavelength responses of $B_{u1}(\lambda)$ and $m_{u1}B_{u1}(\lambda)$ at a wavelength step of 1 nm. Their large variations vs. wavelength are mainly due to the grating couplers for coupling light into the OFD chip, which can be improved with better coupling designs. Similar results can be obtained for $j = 2,3,4$.

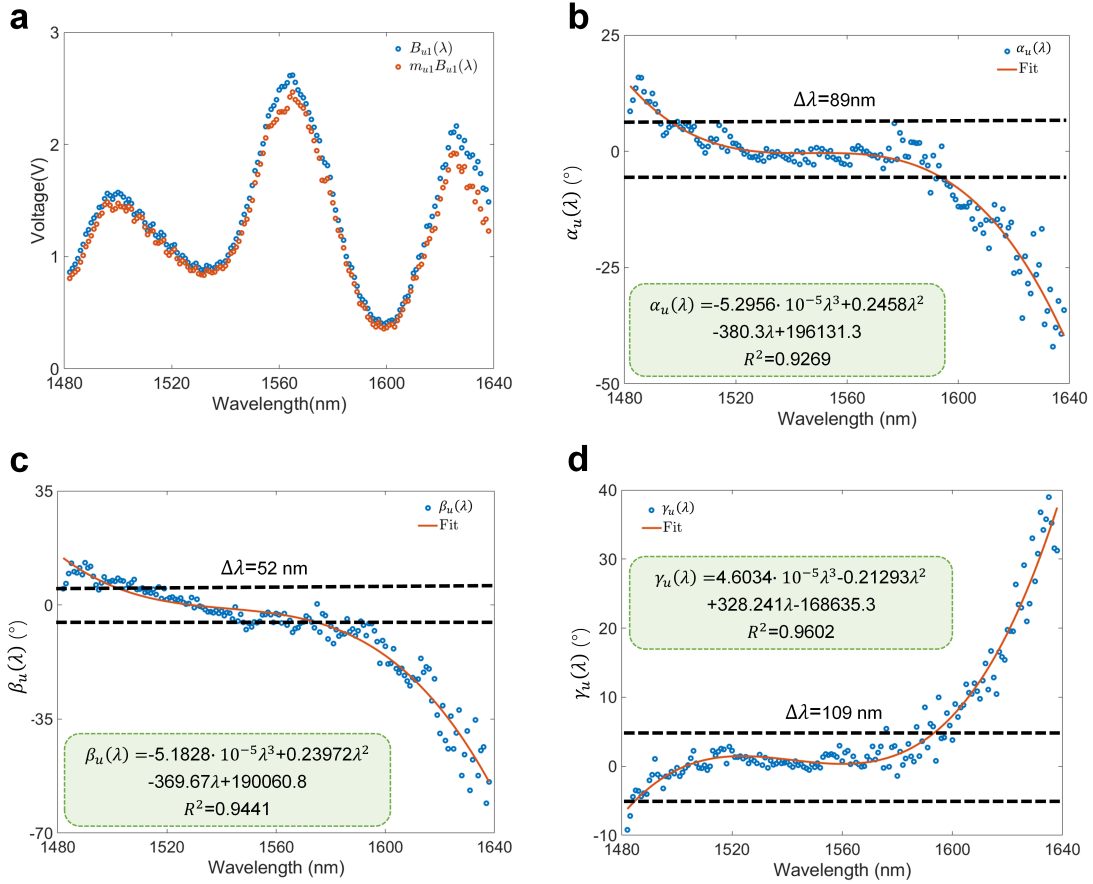

**Fig. S2. Calibration results of the upper interferometer. a** $B_{u1}$ and $m_{u1}B_{u1}$ vs. $\lambda$, **b** $\alpha_u$ vs. $\lambda$, **c** $\beta_u$ vs. $\lambda$, and **d** $\gamma_u$ vs. $\lambda$, all obtained from the elliptical fits of Lissajous figures of $V'_{u1}(t)$ vs. $V'_{u3}(t)$, $V'_{u2}(t)$ vs. $V'_{u3}(t)$, and $V'_{u1}(t)$ vs. $V'_{u4}(t)$ with $\lambda$ at 1 nm increment. A Santec tunable laser (TSL-570) and a DAQ card (ART USB3133A, 16-bit resolution, 60kS/s maximum sampling rate) are used for the calibration.

Figs. S2b, S2c, and S2d are the wavelength responses of phase imperfections $\alpha_u(\lambda)$, $\beta_u(\lambda)$, and $\gamma_u(\lambda)$ obtained from the elliptical fits of the Lissajous figures. As can be seen, they are near zero in the vicinity of the design wavelength 1550 nm and increase significantly beyond 1580 nm. The bandwidths corresponding to ±5° phase deviation for $\alpha_u(\lambda)$, $\beta_u(\lambda)$, and $\gamma_u(\lambda)$ are 89 nm, 52 nm, and 109 nm, respectively, with much large phase deviations outside of these bandwidths. Fortunately, these large imperfections are already accounted for in Eq. (S9) such that they will not impact the accuracies of the $\Delta f(t)$ calculation if their values are determined

by calibration. As discussed previously, these five parameters, together with the circuit bias $A_{uj}$, are stored in the lookup table to be called out when calculating the frequency variation $\Delta f(t)$ using Eq. (S9) if the wavelength is approximately known.

As mentioned in the discussions below Eq. (S6), in addition to the parameters obtained above, the FSR of the upper I-Q interferometer is also wavelength dependent due to the dispersion of the TFLN waveguide, which can be obtained using the setup of Fig. 2 by scanning the wavelength of the tunable laser from 1480 to 1640 nm while recording the four sinusoidal interference signals of the upper interferometer. With some data processing, the FSR at each wavelength is taken as the average periodicity of the four sinusoidal signals, as shown in Fig. S3a, which can be used to obtain the corresponding dispersion coefficient, as shown in Fig. S3b.

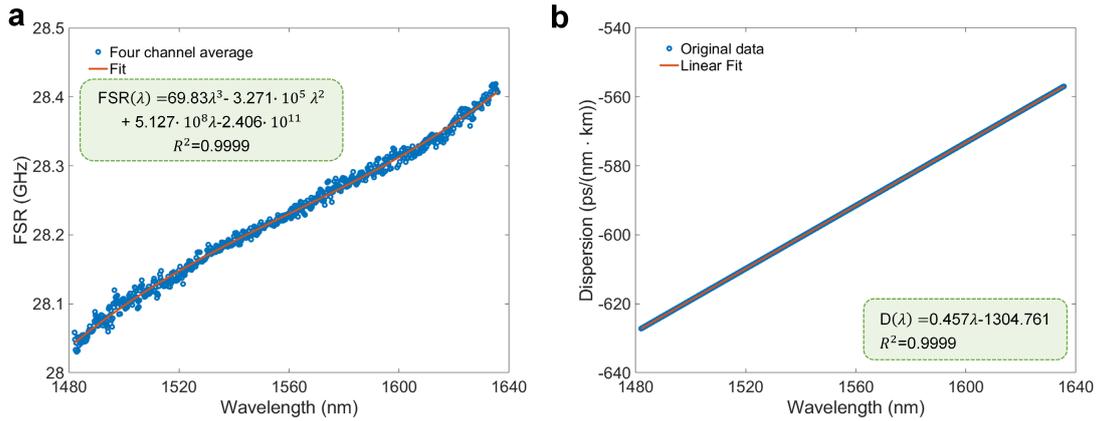

**Fig. S3. Dispersion measurement. a** Measured FSR vs. wavelength. **b** Dispersion coefficient obtained using the data in **a**.

### 3. Lower (assistive) I-Q interferometer calibration

As mentioned in section 2, the lower or the assistive interferometer can be used to determine the absolution wavelength $\lambda$ or frequency of unknown light sources from the lookup table of $\Delta\varphi_l$ vs. $\lambda$ obtained from calibration. Like the calibration of the upper or the main interferometer, the four photocurrent outputs of the lower interferometer with a large FSR ($FSR_l$) are converted into four voltages $V_{lj}(t)$ ($j = 1,2,3,4$) by another 4-channel transimpedance amplifier before being digitized by the DAQ card as the wavelength of the calibration laser is scanned using the setup of Fig. 2. The circuit bias $A_{lj}$ can be obtained by turning the laser off. Because all the components used in the lower interferometer, including the 1×2 coupler and the 2×4 MMI, are identical in design and fabrication to those in the upper or main interferometer, each of the parameters $\alpha_l(\lambda)$, $\beta_l(\lambda)$, $\gamma_l(\lambda)$ in Eq. (S10) should be the same as the corresponding parameter $\alpha_u(\lambda)$, $\beta_u(\lambda)$, $\gamma_u(\lambda)$, respectively. In addition, $B_{lj}(\lambda)$ should have the same wavelength response as $B_{uj}(\lambda)$, except that they may differ from each other by a constant factor which can be determined experimentally. Finally, $m_{lj}$ and $m_{uj}$ are also expected to be the same. Therefore, one may use the same parameters obtained for the upper interferometer in Eq. (S10) to calculate the phase $\Delta\varphi_l(\lambda)$ using $V_{lj}(t)$ ($j = 1,2,3,4$) at different $\lambda$, as shown in Fig. S4a, which can be stored in the lookup table of $\Delta\varphi_l$ vs. $\lambda$ to be called out later when calculating the frequency or wavelength with the upper interferometer using Eq. (S9). For convenience, the corresponding $\lambda$ vs. $\Delta\varphi_l$ can be obtained, as shown in Fig. S4b, which can be fitted to a 4$^{th}$ order polynomial function. Either the lookup table or the fitted function can be used in practice to get the value of $\lambda$ from a measured $\Delta\varphi_l$.

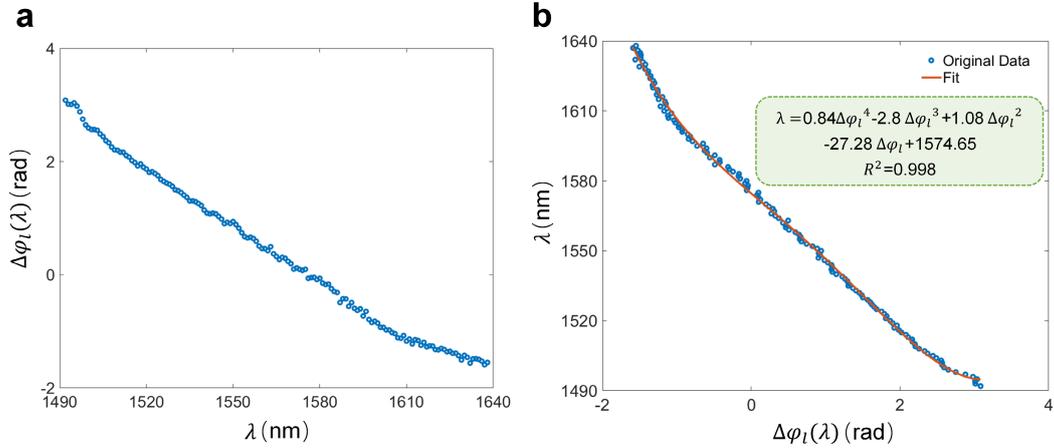

**Fig. S4. Calibration results of the lower interferometer. a** Obtained $\Delta\varphi_l$ as a function of $\lambda$ using Eq. (S10). **b** $\lambda$ as a function of $\Delta\varphi_l$ obtained from the data in **a**. The wavelength step in **a** and **b** is 1 nm.

## 4. Device Fabrication

Figure S5 presents the fabrication process of the proposed device using an x-cut LNOI wafer (NANOLN), which consists of a 360-nm-thick X-cut LN layer and a 4.7-μm-thick buried oxide layer. The fabrication process is detailed below: a Clearing the wafer; b Spin coating the LN film with photoresist; c Patterning the waveguide layer with deep ultraviolet (DUV) lithography; d Etching the LN film; e Removing the photoresist; f depositing SiO$_2$ layer on the surface of the LN film.

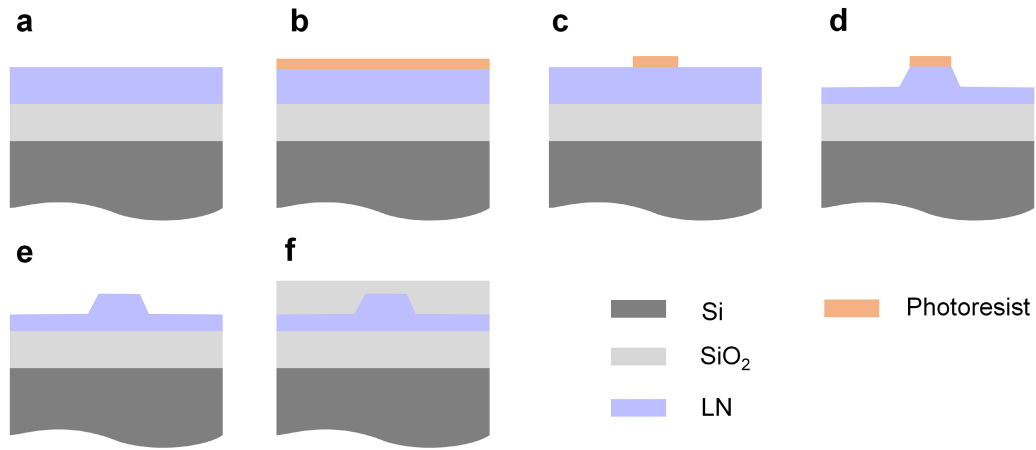

**Fig. S5. The fabrication process of our sine-cosine OFD chip (a-f).**

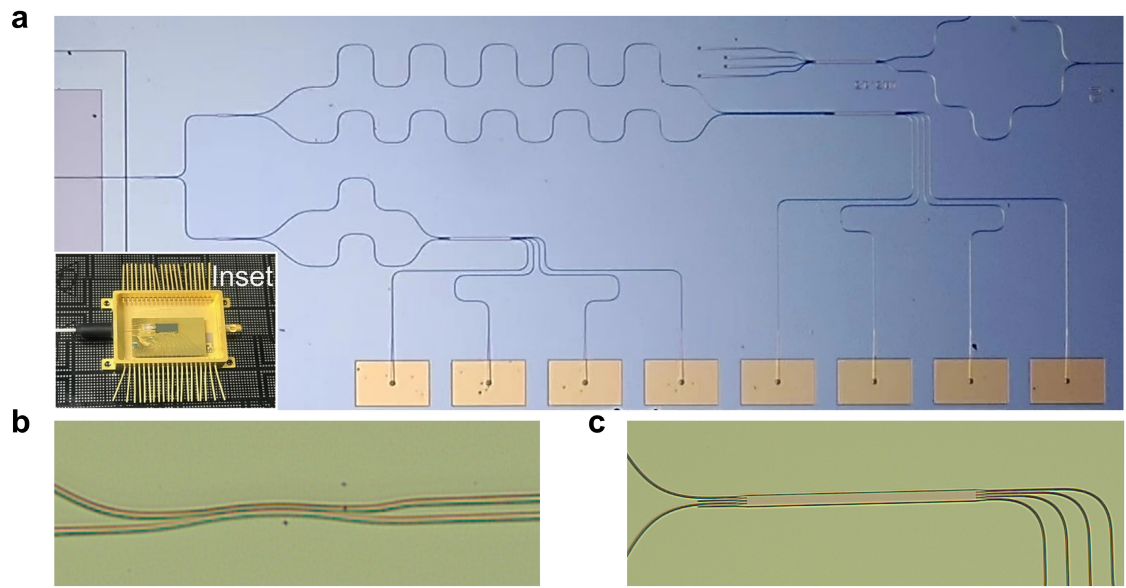

**Fig. S6. Device details.** Micrographs of the TFLN sine-cosine OFD chip **a** consisting of two unbalanced I-Q interferometers, each containing a bend-directional coupler **b**, and a 90° hybrid made with a 2×4 multimode interferometer (MMI) **c**. Inset: the photo of a packaged PIC OFD.

Figure S6 shows a micrograph of the fabricated device. It consists of an edge coupler, three 3dB 2×2 bent directional couplers (BDCs), two 90° hybrids and eight photodetectors to form the upper and lower interferometers. Magnified photomicrographs of a 3-dB 2×2 BDC and a 90° hybrid are shown in Fig.S6b and Fig.S6c. The inset shows the packaged device in a butterfly housing with a thermoelectric cooler (TEC).

**Supplementary References**